\documentclass[12pt]{article}
\usepackage{graphicx}
\usepackage{amssymb,amsmath,amsfonts,palatino,amsthm}
\usepackage{amssymb}
\usepackage{epstopdf}
\newcommand{\f}{\begin{equation}}
\newcommand{\ff}{\end{equation}}

\begin{document}

\title{Views, variety and quantum mechanics}
\author{Lee Smolin\thanks{lsmolin@perimeterinstitute.ca} 
\\
\\
Perimeter Institute for Theoretical Physics,\\
31 Caroline Street North, Waterloo, Ontario N2J 2Y5, Canada\\
and\\
Department of Physics and Astronomy, University of Waterloo\\
and\\
Department of Philosophy, University of Toronto}
\date{\today}
\maketitle

\begin{abstract}

A non-local hidden variables theory for non-relativisitic quantum theory is presented, which gives a realist completion of quantum mechanics, in the sense of a complete description of individual events.    The proposed fundamental theory 
is an extension of an energetic causal set theory, 
(\cite{ECS1}- \cite{ECS4}),  
which 
assumes
that time, events, causal structure, momentum and energy are
fundamental.    But space and the  wave function are emergent. 

The beables of the theory are the {\it views} of the events,
which are a subset of their causal pasts.  Thus, this theory asserts that the universe is a causal network of events, which
consists of partial views of itself as seen by looking backwards from each event.  

The fundamental dynamics is based on an action whose potential energy is proportional to the variety, which is a 
measure of the diversity of the views of the events, while
the kinetic energy is proportional to its rate of change. 
The Schrodinger equation is derived to leading order in an expansion in density of the events of the fundamental histories.
To higher order, there are computable corrections,  non-linear in the wave function, from which  new physical effects may be predicted.


\end{abstract}

\newpage 

\tableofcontents

\section{Introduction}

\begin{quote}

{\it "Inconsistencies challenge the legitmacy of the theories they 
afflict \cite{KS}  "}  

\end{quote}

In this paper we further our understanding of  a class of 
realistic theories which are posited to be satisfactory completions
of both general relativity and quantum theory.

Unlike an interpretation, a completion requires the introduction of new dynamics.   

These theories  are based on a number of ideas, some of which are new here, others of which 
have appeared in papers over the last several 
years\cite{CTV, ECS1,ECS2}.

The belief that quantum-mechanics is incomplete is as old
as the theory itself.   
The first  completion of quantum
mechanics, was proposed, by Louise de Broglie, even 
before quantum mechanics was fully formed\cite{pilot}.
David Bohm rediscovered it 25 years later\cite{Bohm}.  Other 
realist completions of quantum mechanics include
Nelson's stochastic mechanics\cite{Nelson}; meanwhile 
Steve Adler\cite{SAdler}  Artem Starodubtsev\cite{Artem}
and Markopoulou and the author constructed hidden variables theories from matrix models\cite{hidden2002,hidden2003}

In earlier proposals for completions of quantum mechanics, I
introduced the concept of non-local and relational hidden
variables theories\cite{hidden0,hidden}.   These  theories were
developed in a number of 
versions\cite{hidden2002,hidden2003,hidden2006,hidden2007}.

The class of theories we study here are called  the {\it causal theory of views. (CTV)}    They were introduced in \cite{CTV},
and were built on a wider framework called 
{\it energetic causal sets}, which were developed with Marina 
Cortes\cite{ECS1}-\cite{ECS4}.
They share some features with causal set models\cite{cs} - although also some differences.   Other models which embelish 
causal set models with charges and other conserved quantaties were
introduced by Cohl Furey\cite{Cohl} and Fotini Markopoulou\cite{Fotini1}.

The role of variety\cite{variety1}  in the formulation of completions of quantum mechanics was introduced in
\cite{real2} and developed in \cite{CTV}.   Some related ideas
concerned with the real ensemble formulation were
introduced in \cite{real1} and developed in\cite{real2} and
\cite{CTV}.

     \subsection{The main ideas}

\begin{itemize}

\item{} The first idea is that space is emergent, but time in the sense of causality is fundamental\footnote{The first person to proppose this option forcefully was
Fotini Markopoulou\cite{FMst}.}

\item{} But if space is emergent, at the fundamental level there is no space, hence no notion of distance,
hence no notion of locality.    And no notion of non-locality, either.   Locality and non-locality are both emergent.  Hence
so must be the distinction between quantum and classical dynamics.

But dynamics is about moving in space.  Without space, 
there are no distances to measure the fall-off of forces, there are no fields,
no derivatives etc.  What can dynamics involve?

Also, without space, what is a reference system?  What is the purpose of a symmetry?

Einstein had several different motivations in his search for general relativity.  
Rather late in the process, he understood the role of gauge invariance under
active diffeomorphisms.  It took a good think through the hole experiment-which he initially misunderstood.   Once he had that he was essentally done.  But another motivation was to relativize the concept of
inertia so that there was an expansion of the relativity principle from an
equivalence of inertial frames to a  {\it general} principle of 
relativity under which  {\it all} frames would be equivalent.
In this he failed, which is good because the premise is wrong.

There seems to be no equivalence between inertial and accelerating motion.
But if space takes a walk that distinction also  disappears and there is a path to start with a general equivalence of observers.
This is the motivation for what follows.  

\item{} We democratize and universalize the notion of a frame of reference by replacing it with the 
notion of a {\it view}   of  an event \cite{CTV}.

In most experiments we deploy just one or two frames of reference.   But every event has a view, which is short for its view of its causal past.   That is the information available from physical degrees of
freedom whose coincidence constitutes that event, about the past causal progenitors of that event.   

\item{}We then replace measures of {\it distance,} measured according to a fixed metric, $q_{ab}$,
in background dependent formulations of physics, with the measures of {\it differences}
between views.

We postulate that nature has an innate capability of judging differences between views.   This may seem strange, but it is no
stranger than the familiar  metaphysical assumptions that underlie standard, background dependent physics.   
Namely to assume that particles of matter automatically know their distances to all other particles, as is necessary if
the particle is to be able to figure out how much and in what direction it should accelerate.  Or, alternatively, how do particles "know" what representations they are members of and what their couplings should be to each of the other fields.  

\item{}We propose to construct dynamics for a background independent theory with no space to be embedded in,
directly in terms of differences between views.   If we posit that the relations that define individual events are causal 
relations, then the views are views not just of the nearby neighbourhood but of the causal past.

\item{}To some extent, increasing distances may have the effect of increasing a 
difference of views, as new items come into view as you travel
away from your starting point.  But there may also be two atoms or molecules, far away from each other in physical
space, which have very similar views, given that being simple systems they have only a few qubits of information.
These are then both members of an ensemble containing similar views, which includes them and a myriad of other particles.  We will see that,
by virtue of having similar views, they interact strongly with each other.

\item{} Nature can then distinguish unique views, which have no copies anywhere in the universe.

\item{} Unique events (ie events with unique views), have definite values, they are the be-ables.  

\item{}  Unique views, we postulate, have definite causal relations amongst themselves.  This is part of the resolution of the question of whether causal structure is definite.


\item{} We want to express the postulate that the dynamics of the particles in the universe act to increase the total diversity of
their views.  A measure of that diversity is called the 
variety; roughly speaking it is defined as the sum over all pairs of views of a function of the distance between those views.
Increasing diversity is accomplished  by setting the potential energy equal to a constant times the 
variety\cite{variety1,AU}.

\end{itemize}

Our theory is intrinsically relativistic, as must be  any theory whose ontology includes discrete events 
related by causal processes\cite{cs,ECS1,ECS2}.  
Our theory generates causal sets that, when they
embed in an emergent space, embed into a Minkowski
spacetime\cite{cs,ECS1,ECS2}.

But at the dynamical level the theory breaks lorentz invariance 
in that the Hamiltonian consists of a sum of kinetic energy
and potential energy terms, and their is no simple boost
generator that mixes them.

In this paper we discuss mainly the non-relativistic limit,
developing ideas and tools that we hope will be of use in
more realistic theories.  So we will take the non-relativistic
limit earlier  than in the previous presentations of this
idea\cite{real2,CTV}.

 \section{The kinematics of views}
 
 First, we give a reformulation of the causal theory of views, which has been described previously in \cite{CTV}.
 
 \subsection{Kinematics without space }
 
 We begin by defining the components of a causal theory of views.
 
 We posit that an {\it  history} is an {\it energetic  causal set}, $\cal E C S$, which consists of a set  of events, $E= \{  I,J,K,..\} .$ related by a set of direct causal links, $ L_I^J$, 
 Each event $I$ has a set of $n_I^p$ parents
 and a set of $n_I^c$ children.   A single event may have be up to $n_{pre}$ predecesors, and each of these may have up to $n_c$ progeny.
 
 Note that the causal links are a subset of the causal relations, which are inferred  from the causal links by transitivity.
 
To each link, ${}_I^{\ J}$, there is 
 attached an energy-momentum $ p_{\alpha J}^K \in {\cal P} $ that links event $K$ to event $J$.     When the event $J$ occurs, that much  energy-momentum is transferred from event $K$ to event $J$.   
 
 The $P_\alpha $ live in an 
 energy-momentum space  $\cal P$, which has a norm and may have a non-trivial connection.  In the rest of this paper we shall assume that this energy - momentum space is flat. 
The question of how curving the momentum space affects the
physics is closely related to the relative 
locality theories\cite{rl1,rl2}, which indeed was one of the inspirations for ECS\footnote{Notation:  $\alpha =0, 1,2,3$ is a spacetime index,   $a$= 1,2,3 is a spatial index, 
 while $I,J,K = 1 \ldots N$  is a label on the events.}.
  
 At each event, $I$ there is a conservation law for 
energy-momentum.
 This is expressed as:
 \f
 {\cal P}_{a^I} = \sum_{K \in  {\bf IPast}(I) } P_{ a \ K}^I -  \sum_{L \in {\bf IFut}(I)}  
 P_{a I}^L    =0
 \label{P=0}
 \ff

 Here ${\bf IPast}$ and ${\bf IFut } $ are the immediate past and immediate futures of the event $I$.   We also denote this by
 $J |> K$.
 
  The collection of information about the incoming energy and momentum tranferred to an event from an event's
 immediate predecesors is called the {\it view of the event,} 
 ${\cal V}_I$.  The simplest way to represent it is by means
 of an uordered list.
 \f
 {\cal V}_I = \{ p_{\alpha J}^{I} ,   \ldots     \}.
 \ff

\section{Dynamics for theories of views}

\subsection{Taking the non-relativistic limit.}
 
 The energy-momentum transferred, $p_{\alpha K}^J$,
 have dimensions of momentum, 
 so 
 \f
 p_{0 K}^J = \frac{1}{c}  E_{ K}^J 
  \ff
  
  We definie the non-relativistic limit by the limit
  $c \rightarrow \infty $, which implies that all the zero components vanish   $ p_{0 K}^J \rightarrow 0 $.
  Similiarly 
  
  \f
  x_0 = \frac{t}{c}    \rightarrow 0
  \ff
 
\subsection{Half-quantum theory:  the half integral} 
 
 The dynamics is defined by a path integral, but of a rather unusual sort.
 Normally one defines the canonical path integral as an integral over phase space-seen as a bundle over the configuration
 space.
 In many simple cases, the  integrals over momentum are Gaussian and can be done directly.   The result is an
 integral over configuration space.
 
 The theory we are presenting here is structured a bit differently.  One of our core ideas is that the fundamental theory is defined  only in terms of momenta variables plus causal structure. Spacetime is to emerge.
 
 So we begin with what we call a {\it half integral\cite{ECS2}.}
 \f
 \boxed{
 Z[\Gamma ]= 
 \prod_{J|> K \in \Gamma }   \int \ dp_{a \ J }^{K }   \prod_I \delta ( {\cal P}_a^I )
 e^{-\imath{\cal   H } (P )  }}
 \label{action0}
 \ff
 
 We call this {\it half-}quantum theory, defined by the {\it half integral},
  because we integrate over momentum
 and energy but not position, which indeed nowhere appears.
 Since only half of the canonical pair appears, there are no 
 canonical commutation relations.     Hence, if $\hbar$
 appears at some point,  there will have to be a different
 role for it.

 
 We define the half-integral for a specific causal network, denoted
 by $  \Gamma$, so all events, $I$ are in $\Gamma   $.
 We will here focus on these integrals, with $\Gamma$ very large
 and complex, in the sense of halving large values of 
the variety  ${\cal V} [\Gamma ] $.  In this paper
$\Gamma$ is fixed as we are interested in the sums over
labels on $\Gamma$.  We will consider varying $\Gamma$
later.

\subsection{Exponentiating the constraints}
 
 To give the full  definition of the theory (\ref{action0})
we next want to specify the Hamiltonian,  ${\cal H}$.
 
 The first step is to exponentiate the constraints that generate energy-momentum conservation in 
 (\ref{action1}).
We introduce the dual space to the momentum space   
${\cal P}^*$, which has elements $z^a_I$.
We need one for every event.
\f
\boxed{
 Z= 
 \prod_{J |> K} \int \ dp_{a \ J}^K  
 \prod_{I} \int dz_I^a
 e^{-\imath  \sum_I  z_I^a  {\cal P}_a^I  -  {\cal   H } (P )  }}
 \label{action1}
 \ff

Next we specify the form of the  Hamiltonian ${\cal H}(P)$
which we will take to be the sum of kinetic and potential energy
terms.
 \f
 {\cal H} = g T  + g^\prime U
 \ff


Note that he  dimensions of $ [g] = \frac{time}{mass} $, while
 the dimensions of $[  g\prime] = \left [ \frac{\hbar^2 }{8 m g^2} \right ] $.
 
So the effective action is given by
\f
S^{eff} =  \sum_I  z_I^a  {\cal P}_a^I  - 
 g T  -  g^\prime  U
\ff

\subsection{Difference and variety}
 
The next step is to construct detailed kinetic and potential energy
 terms.   Both are based on the idea of variety, which is a measure of how diverse the views are.  The kinetic energy, 
${\bf T}$, measures the rate of change of views
 among causally related events.  $U$ is a potential energy that measures
 the diversity of views among causally unrelated events.
 
 These are both based on measures of difference between pairs of views,
 which is generically denoted ${\cal D}(I,J)$.   The sum over these gives
 the total variety:
 \f
 {\cal V} =\frac{2}{N(N-1)} \sum_{I,J} {\cal D}(I,J)
 \ff
  
  A convenient form for ${\cal D}(I,J)$ is
  \f
  {\cal D}(I,J) = (  {\cal W}_I -    {\cal W}_J    )^2
  \ff
  
  where ${\cal W}_I$ is the view seen by or at the event, $I$.
  
The different theories in this class come from three choices which must be
  made. 
  
  \begin{enumerate}
  
 \item{}how the views are defined and represented,
  
  \item{}  how the  the differences $D(I,J)$ between views are
defined and 

\item{} which pairs of views are summed over to define
the kinetic and potential energies.

  \end{enumerate}
  
  In the theories discussed here,  we take the views to be a set of incoming energy momentum vectors, generally density weighted, where the weight is $w$.
\f
{\cal W}_{a}^{ I} = 
 \sum_{K \in {\bf Past (I) }} 
\frac{p_{a K}^{I}  }{| p_{a K}^{I}  |^w }
\ff


 I believe it may be the case that there is a single universality class (or just a few)
 among these theories, but at the present state of knowledge  I look for simple
 choices that lead to simple derivations of the recovery of quantum mechanics.

 \subsection{Constructing the kinetic energy}
 
 Consider an event $I$ and an event $J$ in its immediate causal past.  We indicate this by
 \f
 I |> J
 \ff
 We define the {\it surprise} at an event $I$ to be a measure of
 how much  it differs from its immediate predecessors.
 \f
 {\cal S}urprise (I)= |  
 \sum_{K \in {\bf IPast} (I)} {\cal D}(I,K)  |^2
 \ff

So one simple definition of kinetic energy is
\f
{\cal T}_{surprise} = \sum_I {\cal S}urprise (I) 
\ff

Here we will make use of a similar form.

We define the causal variety with $N$ events by
 \f
T
 = 
 ( \sum_{ I |> J}  (  {\cal W}^{(p=0)  \ I}_a  -   {\cal W}^{(p=0)  \ J}_a      )^2
\label{KE}
\ff

Written out this is
\f
 T
= 
( \sum_{ I |> J}  \left (  \sum_{ I > K}  p_{aK}^I -
\sum_{ J  > L}    p_{a L}^J    \right )^2 
\ff

\subsection{Constructing  the potential energy}

The potential energy is
\f
{\cal U} = \sum_{I<>J} {\cal D}(I,J) _{w=2}
\ff
where $I<>J$ means $I$ and $J$ have no causal relation, ie analogous to a spacelike separation.

Written out this becomes

\f
{\cal U} =   {\bf U}_{<>}
= 
 \sum_{ I <> J}  \left (  \sum_{ I > K}  
    \frac{p_{aK}^I }{|p_{aK}^I |^2} -
\sum_{ J  > L}   \frac{p_{aL}^J }{|p_{a L}^J |^2}  \right  )^2 
\label{pote}
\ff

\subsection{Summary of the theory}

 We define the Hamiltonian:
 \f
\boxed{ {\cal H}  =  g T  + g^\prime{ \cal U}}  
 \label{Seff}
 \ff
 Note that the only geometry referenced in the action is the geometry of
 momentum space ${\cal P}$. 
 
 In the course of deriving the Schrodinger equation to
 leading order in $g'$,  we will find also that to zeroth order in
 $g'$, the Hamilton Jacobi equations are satisfied.  
 This implies the conservation in time of the Hamiltonian,
 (\ref{Seff}).

The corresponding half integral is

\f
\boxed{  Z= 
\prod_{J |> K}   \int \ dp_{a \ J}^K  
\prod_{I}  \delta ( {\cal P}_a^I )   
 e^{\imath (gT  + g^\prime {\cal U} )} }
 \label{action-half}
 \ff
 
This defines the theory\footnote{It is easy to check that the dimensions of $g$ are
$ \frac{time}{mass} $ while the dimensions of
$\frac{g^\prime }{g^2}  $ are  
those of
$ \frac{\hbar^2 }{m} $.   Note that we do {\it not}
use units in which $\hbar$ or $c$ are equal to one.}.

In fact, we will see as we go along that we will want
to choose
\f
\frac{g'}{g^2} =  \frac{ \hbar^2 }{8 m}   Z_V
\ff
to get quantum mechanics to emerge at the end.
Indeed, it is only through this choice for $g'$ that
$\hbar$ enters the theory at all.
 $Z_V$ is a dimensionless constant, defined below.

\section{Evaluating the half integral}

What are we going to do with this theory?   There is no space
or spacetime.   The only time ordering is causal order.

How do we get physics out of it?
We start our anylysis with the form resulting from exponentiating the constraints.

\f
\boxed{  Z= 
\prod_J^K   \int \ dp_{a \ J}^K  
  \prod_{I} \int dz_I^a
 \ e^{\imath (  - \sum_I  z_I^a  {\cal P}_a^I +
g T + g^\prime {\cal U}) } }
 \label{action-half3}
 \ff



The above expression, with the effective action
\f
\boxed{
S^{eff} =  - \sum_I  z_I^a  {\cal P}_a^I +
 g T + g^\prime {\cal U}} 
 \label{action17}
\ff

 gives us a somewhat different, but equally precise definition for
the theory; we then proceed to evaluate the partition function.
We do this in a series of steps.

We have picked the order of integration to integrate over the
momentum variables, $p_{a M}^N$ first, for fixed values
of the $z^a_I$.   Then at the end we will integrate over the
$z^a_I$ in a certain course grained approximation.

\subsection{Observbles}

Let ${\cal A} [z^a_I ]$  be a functional of the $N$ $z^a_I$'s,
as well as the choice of an $N$-event causal set.
Then the expectation value is defined by
\begin{eqnarray}
<  {\cal  A } (z^a_I ) >   &= & \frac{1}{Z} 
\prod_J^K   \int \ dp_{a \ J}^K  
  \prod_{I} \int dz_I^a
{ \cal A} (z^a_I )
 e^{\imath S^{eff} }
 \nonumber
 \\
  &= & \frac{1}{Z}  
  \prod_{I} \int dz_I^a
 { \cal A} (z^a_I )   \prod_J^K   \int \ dp_{a \ J}^K  
 e^{\imath S^{eff} }
 \label{action-half3}
 \end{eqnarray}
 
 where we have exchanged the order of integration.

\subsection{Integrating over the transverse momenta}


We are going to approximate these integrals by the
stationary phase approximation.   To do this it will be
convenient  to divide the momentum fluctuations into
transverse ad longitudinal parts and then begin  with the 
former.

The result of integrating over the momenta is very simple, so I state it up front. 
We  will be able to satisfy the condition of
stationary phase for $p_{a K}^{L}$ when
$z^a_K$ and $z^a_L$
are located so that
\f
p_{a K}^{L} \ g^{ab}  = \frac{1}{g n_{pre}} [   z_L^b - z_K^b  ]  
+ {\cal O}(g'^2 ) .
\ff
here $g^{ab}$ is the metric on momentum space.
That is going to tie down the momenta transferred 
from event $K$ to event $L$ 
to be
a function of the differences between the 
two $z^a_K$ of the events they connect.  

Recalling that $g$ is proportional to a time per mass,
this is just the ordinary,
\f
p = m \frac{\Delta z}{\Delta t}
\ff

Recall that there is one of these relations for every
pair of connected events.  Each can be thought of as a little fragment of a three dimensional space.   But to assert that space  has emerged we have to find consistent solutions for
all $J |> K$.   

\begin{itemize}

\item{}{{\it Split the integrals over momentum transferred, }
$p^I_{a J}$,  for each
pair of causally connected events, $I |> J$,
into transverse and longitudinal  fluctuations.}


In the neighbourhood  around a particular $p_{a, J}^{K}$
(technically the cotangent space of $\cal P$) there
are two kinds of fluctuations (or small variations): transverse 
and
longitudinal.  These  may be defined in terms of
 projection operators:

 \f
 {\cal T}_{ac} = [\delta_{ac} - \frac{ p_{ aK} ^L   p_{ cK} ^L  }{|  p_{ aK} ^L |^2} ] , 
 \label{PT}  
 \ff
 
 \f
 {\cal L}_{ac} =  \frac{ p_{ a J} ^{N K}  p_{ c J} ^K  }{|  p_{ a J} ^K |^2} 
 \label{PL}
 \ff
 \f
  {\cal L}_{ac} +  {\cal T}_{ac} =  \delta_{ac}
 \ff
 independently for each $J$ and $K$.

 We can then, for each $J<|K$    divide the fluctuations in
 the momenta $\delta p_{a J}^K$  into transverse
 and longitudinal modes 
\f
\delta p_{a J}^{L K} =  {\cal L}_{a}^{\ c} \delta p_{cJ}^{L K},
\ \ \ \ 
\delta p_{a J}^{T K} =  {\cal T}_{a}^{\ c} \delta p_{cJ}^{T K}
\ff
We will write
\f
p_{a L}^{\ \ M}  =  l_{a L}^{\ \ M} +  t_{a L}^{\ \ M} 
\ff

\item{}{ Find equations of motion for
transverse momentum fluctuations.}

We are going to carry out the integrals over the transverse
momenta first.   We make use of the stationary phase approximation,
which means we have to work out the equations of motion
for the transverse momentum modes.
 \f
  \frac{\delta S^{Seff} }{\delta t_{c M}^{\ N}}  =0
 \ff

Each of the three terms in the effective action
contributes to the fluctuations of the momentum modes.

\item{}{\it Compute the gradiant of the potential energy.}
 
We first  compute the term from the gradient of the potential energy
\begin{eqnarray}
\Pi^{c J }_K  &=& \frac{\delta U }{\delta p_{c K}^J}
\nonumber
\\
&=& 2 \frac{1}{{| p_{ a K}^K  |}}
 {\cal T}^{c d} 
( {\cal W}_d ^J - {\cal W}_d^K    )
\label{dV}
\end{eqnarray}

 So we find that $U$ is pure transverse.  Of course might
 have known this from the fact that it is composed of
 transverse functions,  
 $\tilde{p}_{a J}^K = \frac{{p}_{a J}^K}{| {p}_{a J}^K        |^2}.$ 

\item{}{\it Compute the gradient of the kinetic energy}


For the kinetic energy it is easiest to choose 
the density weight 
$w=0$.   Together with the fact that
we only compare views one causal step apart, it is much simpler to evaluate.
It mostly comes down to separating cases.

Consider an event $I$ and its view, which is made up of the events in its immediate causal past.

Notice that there can be multiple paths between two events,
whih implies that there may be closed paths on the 
momentum space.
For example, 
suppose $I$ has two direct antecedents, which are $J$ and $K$.     Lets further
consider $L$ which is an antecedent of both $I$ and $J$, but not $K$.

There are two paths to get from $L$ to $I$, one direct via the link $
p_{a L}^I$.   Or the energy may travel from $L$ to $J$ via $p_{a L}^J$
and then From $J$ to $I$ on  $p_{a J}^I$.

We posit that two momentum vectors are added together by parallel transporting the first along the second.  Then they must satisfy the linear equations.
\f
p_{a L}^I =p_{a L}^J + p_{a J}^I
\ff
Now the kinetic energy is proportional to square of the difference between ${\cal W}_I$
and ${\cal W}_J$

More precisely 
\f
T= 
\sum_I ( \sum_{J \in {\bf Past}(I)}  p_{a I}^J -    \sum_{K< {\bf Past} (J) }  
p_{a K}^J )^2
\ff
It is easy to see in this case this is just
\f
T=  \sum_{I |> J}    \frac{D_I^2}{2} ( p_{a I}^J )^2 
= \sum_{I |> J}    \frac{D_I}{2}  \left  (  ( t_{a I}^J )^2 
+ ( l_{a I}^J )^2 \right ) 
\ff

So the contribution to the propagation or equation of motion of the momenta from
the kinetic energy is both longitudinal and transverse

Here $D_I$ is the number of immediate
predessors event $I$ has.  Sometimes this is set
\f
D_I = n_{pre}
\ff 

We can then put all three terms the variation of the action  together to find

\f
\boxed{
p^J_{a K}   =  \frac{1}{ g n_{pre}^2 }
(  z^a_J  -  z^a_K)   -  \frac{g'}{ g n_{pre}^2 }
 {\cal T}^{ab} ( {\cal W}_b^J - {\cal W}_b^K    )}
\label{3together}
\ff

At a stationary phase point, $\Pi_{a M}^N =0 $,  we can project the dual vector on momentum space into longitudinal
and transverse modes with respect to each $p_{a N}^M$.

The result is
\f
(     z^a_M -z^a_M  )_L  =   g n_{pre}^2 l_{a M}^N   
\label{long}
\ff
We also have
\f
(     z^a_N -z^a_M  )_T  =   g n_{pre}^2 t_{a M}^N +g' {\cal T}_{a M}^{N}   
( {\cal W}_b^N - {\cal W}_b^M    )
\label{trans}
\ff

Note that this inversion from a functional of the $p_{a J}^K$ to functional of the $z^a_I$ always exists because the dynamics  ensures
that there is no ambiguity coming from the possibility that the map between
the cotangent spaces of $V$ and and the tangent spaces.   But there cannot
be because such a singular configuration would have infinite energy.

\item{}{  Integration of the potential
energy term.}

We use the information we now have about the stationary
phase approximation to integrate out the transverse
small fluctuations.  The present  situation is
\begin{eqnarray}
 Z &= &  
  \prod_{I} \int dz_I^a
  \prod_J^K   \int \ dl_{a \ J}^K  
  \prod_J^K   \int \ dt_{a \ J}^K  
 \ e^{\imath (  - \sum_I  z_I^a  {\cal P}_a^I +
 g n_p^2 (t^2 + l^2 )   + g^\prime {\cal U}  }
 \nonumber
 \\
 &=  & 
  \prod_{I} \int dz_I^a
  \prod_J^K   \int \ dl_{a \ J}^K  
  {\cal Y}(z^a_I , l_{a K}^J  )
 \label{action-half4}
 \end{eqnarray}

We start with (\ref{action-half4}.)
 We are next
going to do the integral over $\int dt_{aJ}^{\ K}$.

This is of the form
\f
{\cal Y} = 
 \prod_{  J  |>   K}   \int \ dt_{a \ K}^{J} 
 \ e^{ { \imath} \left  (  z_J^a - z_K^a  )( t_a^{JK}  + l_l^{JK} )  -
  g n_{pre}^2  \{ (t_{a K}^{\ J})^2  +  (l_{a K}^{\ J})^2 \} +  g'  U(p_{a M}^{N})   \right  )      }
 \ff
If we set $g'=0$ the integrals simplify to 
pure Gaussian integrals.
\f
{\cal Y}_{g'=0} = 
 \prod_{  J  |>   K}   \int \ dt_{a \ K}^{J} 
 \ e^{ { \imath} \left  (  z_J^a - z_K^a  )( t_a^{JK}  + l_l^{JK} )  -
  g n_{pre}^2  \{ (t_{a K}^{\ J})^2  +  (l_{a K}^{\ J})^2 \} )   \right  )  } 
  \ff
  We perform the integral over the $t_{aJ}^K$, yielding
 
 \f
 {\cal Y}_{g'=0}   \approx   e^{\imath           
 \left ( ( z_J^a - z_K^a  ) l_a^{JK}   +
  \frac{1}{g^2} (z_J^a - z_K^a  )^2
\right ) }   
 \ff

The stationary phase condition for the transverse integral  is
\f
t_{a  J}^{K}= \frac{1}{2 g n_{pre}}  
\left (    z^a_J   - z^a_K
\right )   + {\cal O}(g^{' 2} )
\label{sp}
\ff
Note that the whole of $(    z^a_J   - z^a_K )  $ is involved in
the stationary phase condition (\ref{sp}).

We can then write, an expansion in powers of $g'$

\begin{eqnarray}
{\cal Y}    &    \approx  &   
 \prod_{  J  |>   K}   \int \ dt_{a \ K}^{J} 
 \ e^{{ \imath}     ( (z_J^a - z_K^a  ) t_a^{JK}    -
  g    (t_{a K}^{\ J})^2  
     }    [ 1 +  g'  U(p_{a M}^{N}  ) ]  
     \nonumber
     \\
     & \approx   &
       e^{\imath  \left ( ( z_J^a - z_K^a  ) l_a^{JK}   +
  \frac{1}{g^2} (z_J^a - z_K^a  )^2
\right ) } [ 1 + g'    < U>    +   {\cal O}(g^{' 2} )
\label{Yint}
 \end{eqnarray}

 The potential energy is then a function of the
 $z^a_I$.
 
\f
{\cal U} (z) = < U > = 
\frac{1}{g^2}
\sum_{I<>J}   
\left (   \sum_{K \in {\bf IPast (I) }} \frac{z^a_I -z^a_K  }{| z^a_I -z^a_K   |^2  }-
\sum_{L \in {\bf IPast (J) }} \frac{z^a_J -z^a_L  }{| z^a_J -z^a_L   | ^{2} }  \right )^2
\label{PE(P)}
\ff
 
 The effective action is then a functional of the $z^a_I$ and
 the longitudinal part of the momentum variables
 \f
 S^{eff} (z^a_I ,  l_{a}) =
  \frac{1}{g^2} \sum_{J K}(z_J^a - z_K^a  )^2
 + \sum_I z^a_I {\cal L}_a^I  
 + g n_{pre}^2  g    [ l_{a J}^{K}   ]^2   + g' U^{eff}
 \ff
 where
 the effective potential is, 
  \f
 U^{eff}=  
  \frac{g'}{g^2 }  
  \sum_{I<>J}   
\left (   \sum_{K \in {\bf Past (I) }} \frac{z^a_I -z^a_K  }{| z^a_I -z^a_K   |^2  } -
\sum_{N \in {\bf Past (J) }} \frac{z^a_J -z^a_N  }{| z^a_J -z^a_N   | ^{2} }  \right )^2   +  C + {\cal O}(g^{' 2}  )
\label{PE(P)}
\ff

Here we have defined the constant,
\f
C=\sum_{JK}  \frac{g'}{g^2} (z_J^a - z_K^a  )^2  
\ff
This is a divergent constant and we will remove it in what
follows.

To summarize,  we have carried out the integration over
the transverse parts of the momentum variables.   We
will proceed to carry out integration of the longitudinal
momentum variables, saving the integration over
the embedding coordinates, $z^a_I$ for last.

 \subsection{Integration over the longitudinal momenta}
 
 \item{}{ We are left with integrals to do in the $z^a_I$ and the
longitudinal momenta, $l_{a J}^{\ K}$. }  We have
\f
Z   =   \prod_{I} \int dz_I^a
 \prod_{J \>K}   \int ( \ dl_{a \ J}^K  )_{L}
 e^{-\imath [ \sum_I  z_I^a  {\cal L}_a^I  
 - g D ( l_{a I}^J )^2 
+   g^\prime  U(l ) ] }    
\label{almostZ}
 \ff
 
 where the  potential energy  (\ref{PE(P)} ) is now a functional of the 
$z$'s.

\item{} { Integrating by coarse graining.}

We now want to integrate over the embedding coordinates,
$z^a_I$ and the longitudinal parts of the
moentum transfered, $l_{a J}^K$.   We will do these
integrals by coarse graining.  That is, we are adopting a
Wilsonian point of view over which the path integral over short scale fluctuations, with an infrared cutoff, $L$,
 is equivalent to doing a coarse graining which replaces
 sums over microscopic variables by averages
 over coarse grained collective
 degrees of freedom.
 
In the present case that means we replace the sums over the positions of the
 embedding coordinates, $z^a_K$  by an integral over  probability distributions, $\rho (z )$.   We assume that the number of events in the causal network, which we denote by $N$,   is large.   We emphasize again that these $N$ events and their causal relations become a single quantum particle under
 coarse graining.
 
The details of how a single classical (in this case) relativistic 
particle emerge from a causal network of events, are in the
first two papers on energetic causal sets\cite{ECS1,ECS2}. 
 
 For example, consider an observable ${\cal A}(z^a ) $ which is a function of the $z^a_I$'s.  The basic definition of the
 average or expectation value of ${\cal A}$ is
 \f
< {\cal A}  > = \frac{1}{N} \sum_{K=1}^N {\cal A}  (z_K ) 
\ff
where $N$ is the total number of events in $\Gamma$.  

We will use $\rho (z)$ to replace this with
 \f
< {\cal A}  > = \frac{1}{N} \sum_{K=1}^N 
{\cal A}  (z_K )  \rightarrow \int_{a}^{R}  d^d z \rho (z)  {\cal A}  (z)
\label{basic}
\ff
In the limit $N \rightarrow \infty$ this defines the probability density for configurations, $\rho (z)$.

Let's pause and make a few observations.   First, 
we are here always discussing ensembles of events or
of particles at a given, fixed time, $t$.

Notice also that the precise matching between the microscopic and macroscopic degrees of freedom depends critically on the boundary conditions which are assumed.   The manifold $\Sigma$  with metric $g^{\mu \nu}$ describes the configuration space $\cal C$ of the system we are describing.

If we ask that $< I > =1$ this implies that
\f
\int_{\cal C} \rho = 1
\ff
so $\rho (z)$ is a normalized probability density, for finding the
position of a single quantum particle.  We will
see shortly that it is conserved.

We could impose
instead
\f
\int_{\cal C} \rho = M
\ff
in which case the same probability distribution functions
can describe $N$ non-ingteracting particles.

This is an  important point.  The elementary events are part of a microscopic description.  At this level there are no particles. and also no waves. (or wave functions.)    A large number of
elementary event  correspond to a single (quantum) particle.

More precisely, $z^a$ are coordinates on a configuration space,
which might be  three dimensional space for a single particle or a $M= 3 N$ dimensional space for a system of $M$ 
particles.

Similarly we turn the sums on $I$ to an integral
\f
\frac{1}{N} \sum_i  \phi  (x_{k+i}, x_k )  \rightarrow Z \int_a^R  d^d x \rho (z+x) \phi (z+x,z)
\ff
The possibility that the integral only approximates the sum for finite $N$, because of the roughness of the estimate for the limits on the integral,  is accounted for by an adjustable normalization factor $Z$.   Note that to complete the definition of the
coarse graining we need to choose the limits of
integration.   More on this below.

The use of probability here suggests that we are introducing an
ensemble of similar systems  to represent the probabilities
for the distribution of events in the single system we are studying.  Depending on the context, this could be an imagined ensemble, as in the case of much equilibrium statistical physics,
an ensemble consisting of many similar classical 
universes\cite{MCU}, or a {\it real ensemble} of many similar regions within our universe, as I suggested in several earlier papers ( \cite{real1,real2}.  )

For the purposes of showing the emergence of quantum mechanics from a realistic theory, any of these can work as part of a background story that is put forward to connect these equations to nature.  So we will not touch these issues till the
summary chapter.

\item{}{ \it  Introduce coarse grained observables.}

Before we can represent the final $dz^a_I$ integrals,
we have first to integrate over the longitudinal momenta,
$l_{a M}^N $.   

How do we coarse grain the momentum variables?   
We define a coarse grained function,  which we can think of as
a kind of momentum-energy current, $ p_a (z)$  on $z^a$ such that
\f
I  = \int d{p}_a    \ \ 
\delta \left ( p_a     -    \sum_{ K >L}  [   {p}_a  (z^a_K)  -     p_{a K}^{L}     ]   \right )
 \ff

If we insert this in the integral (\ref{almostZ} ) and integrate
over the remaining longitudinal modes, $d t_{a J}^K$
we will find only functionals of $\rho (z) $ and $p (z)_a$.

\item{}{ \it  Integrating over the longitudinal 
fluctuations:}

That $p_a$ is longitudinal means that 
\f
\epsilon^{ijk} \partial_j p_k = 0
\label{long}
\ff
The solution to which  is that there exists
a scalar phase $S (z) $ such that,
\f
\partial_a S = {p}_a
\ff
 so we express this as 
\f
I  = \int d S(z)  \delta       \left (   \partial_a S - {p}_a
  \right )
 \ff
Again, we insert thos form of $I$ into the integrand, integrate
over the $p_a (z)$ and we end up with a functional
integral
over the density $\rho$ and the $S$.   The result is an
effective action that depends only on the emergent
large scale collective degrees of freedom, $\rho (z) $
and $S(z) $.   These are not only emergent degrees of freedom,
they are functions of the $z^a$ which are themselves
emergent degrees of freedom.

\item{}{    {\it The terms from the conservation of
momentum}}

We study lastly  the constraint term that governs energy momentum conservation,
\f
{\cal X} =  <  \sum_I   P^I_a z^a_I  >  
=  <  \sum_{I |> J }  p_{a J}^I ( z^a_{I} - z^a_{J} ) >
\ff

We can consider, first the classical limit, in which we insert infinite numbers
of events into the trajectory\cite{ECS1,ECS2},
\f
{\cal X} \rightarrow  \ \sum_i \int_{\gamma^i}  ds \  p(s)_a \dot{z}^a 
\ff

Instead we will assume that the events are spread with a density limited by
(), so we find instead
\f
\sum_I \  P_a^I ( z^a_{I+1} - z^a_{I} ) \rightarrow  \int d^dz \rho (z) \dot{z}^a \partial_a S
\ff

We now use current conservation,
\f
 \dot{\rho} = -\partial_a V^a
 \ff
 where $V^a = \rho \dot{z}^a $
 to find  that
 \f
{\cal X}   = \int_{\cal C} {\rho} \dot{S}
\ff

\subsection{ Assembling the pieces}

\item{}{\it   Putting together all these pieces, 
the action becomes simply,}
\f
{\cal S}^{cg} (\rho (z ), S(z)  )  = \int dt \int_{\cal C}  \rho (z,t) 
\left [  \dot{S} +   
g   n_{pre}^2   g^{ab} (\partial_a S (z,t)) (\partial_b S (z,t) 
+ \frac{g^\prime}{g^2}    {\cal   U} (z)      
\right ]
\label{HJ1}
\ff

At the end we have an ordinary path integral  expressed in terms of the
emergent variables, $\rho (z) $ and  $S  (z)  $.
\f
Z(\rho, S) = \int  d  \rho  d S  
e^{\imath {\cal  S}^{cg}   }
\label{ordinary}   .
\ff

Let us stop and have a look at the action, (\ref{HJ1}).

Treating it as a classical action (or just looking at the
stationary phase approximation, we can take the variation
by $\rho$.   This yields immediately the Hamilton-Jacobi
equation plus one new term proportional to $g^\prime.$
\f
\dot{S} +   
g   n_{pre}^2   g^{ab} (\partial_a S (z,t)) (\partial_b S (z,t) 
+ \frac{g^\prime}{g^2}   {\cal   U}  (z)
\ff
We will shortly see that to correspond with the Schrodinger
we have to set
\f
\frac{g^\prime}{g^2} = \frac{\hbar^2 }{8m}  Z_V
\ff
Where $Z_V$ is a constant which absorbs various factors.

The variation of ${\cal S}$ by $S$ yields 
probability conservation.
\f
\dot{\rho} = - \partial_a (  \rho   g^{ab} \partial_b S )  
\label{pc}
\ff

Thus, the action (\ref{HJ1}) has the Schrodinger equation
as its classical equations of motion.

Considering it an action for classical physics, the action
(\ref{HJ1}) has an intriguing structure.
Notice that independent of the presence or absence of
the potential energy proportional to $\hbar^2$,   the
probability density, $\rho $  is conjugate to the phase
variable $S$.     If we were to start with a standard canonical analysis, 
one finds that the non-vanishing Poisson bracket is.
\f
  [   S (x),  \rho ( y )  ] = \delta^n  (x, y )
 \label{uncan}
\ff

\item{} {\it  Normalizations, limits and cutoffs}

There is one last detail which is to specify the limits on the
remaining integrals to ensure probability conservation.

That is, for finite $N$ and finite densities of events,
the $\rho (z)$ has upper and lower limits that reflect the
fact that the integral over $\rho$ stands for a sum
over a finite set, which consists of $N$ microscopic  events, whose coarse description is a single quantum particle.

\item{}{\it   The uv cutoff}

Note that  we have to be careful to impose limits on the integral to avoid unphysical divergences in $  \frac{1}{x}$.  
These divergences are unphysical because for finite $N$ two configuration variables, $x^a_k$ and $x_j^a$, cannot come closer than a limit which varies inversely with the density at $x^a_k$ and $N$.  This is because if $x_k^a$ and $x_j^a$ are nearest neighbours in the distribution, the density at one of their locations is related to their separation.
\f
\rho ( x_k^a ) \approx \frac{1}{N |x_k - x_j |^d }
\ff
where $d$ is the dimension of the cofiguration space.
Hence, for finite $N$ they are very unlikely to coincide. When we approximate the sums by integrals, the integrals representing intervals between configurations must then be cut off by a short distance cutoff $a$ that scales inversely like a power of $N \rho (z)$.
The short distance cutoff $a (z)$ on the integral above in $d^d x$ then expresses this fact that there is a limit to 
$ \frac{1}{x}$  related to the density.   Hence the short distance cutoff is at
\f
a(z) =\frac{1}{( N \rho (z))^{\frac{1}{d}}}
\label{acutoff}
\ff

\item{}{  \it The infrared cutoff}

There is also an infrared cutoff, $R$.
 This tells us that two systems further than $R$ in configuration space do not figure in each other's views.   A key question turns out to be how the physical cutoff scales
with $N$. 
to ensure that the large $N$ limit is uniform, we want to scale $R$ the same
way as $a$.    Certainly the density $\rho$ is real and well defined, so when we
define the uv cutoff by (\ref{acutoff} )   we also have to scale the $ir$ 
cutoff $R$, so that the ratio
\f
L= \frac{R}{a}
\ff
is fixed.   So the long distance coordinate cutoff $R$ must scale like
\f
L= \frac{R}{( N )^{\frac{1}{d}}}
\ff

 We will define
\f
L= \frac{R}{( N )^{\frac{1}{d}}}
\ff
to represent a fixed physical lengths scale which is held fixed when we take the limit of large $N$ at the end of the calculation.  That way, the physical ultraviolet and infrared cutoffs scale the same way with $N$.  But the large scale, infrared cutoff, $L$ can't know about the value of the probability distribution at some far off point $z$, so while $a$ scales with $\rho$, $r^\prime$ doesn't.

As a result when we scale $x$ and $d^d x$ with $a$ to make the integrals dimensionless, we define $r$, such that,
$R=ar$. But we then hold fixed 
\f
L=\frac{1}{\rho^{\frac{1}{d}}} r = N^{\frac{1}{d}}R  
\ff
as we take $N$ large.  
$L$, unlike $r$, is a length. We shall see that $L$  defines a new physical length scale at which the linearity of quantum mechanics gives way to a non-linear theory.

\subsection{Emergence of the Bohmian potential}

Once we get the scalings and the limits right, the
potential energy term (\ref{PE(P)}) can be expanded in powers of 
$\frac{1}{N}^{\frac{2}{d}} $.  The leading term reproduces David Bohm's
quantum potential[].


The continuum approximation to the spacelike or
acausal variety  is,
\f
 {\cal V}  =   \int d^d z \rho (z) Z_V  \int_a^R  d^d x \int_a^R  d^d y
 [ (  
\frac{x^a}{x^2 } - \frac{y^a }{y^2}  )^2   \rho (z+x ) \rho (z+ y ) 
\ff

We do a scale transformation by writing
$x^a =a \alpha^a $ and $y^a =a \beta^a $.  
To get a single integral over a local function we can expand
\f
\rho (z+a \alpha) = \rho (z) + a \alpha^a  \partial_a \rho (z) + \frac{1}{2} a^{ 2} \alpha^a \alpha^b \partial^2_{ab} \rho (z) +  \ldots
\ff
and similarly for $\rho (z+ a \beta) $ and perform the integrations, holding the upper limit $r^\prime$ fixed. 

The  normalization factor  is
\f
Z_V=\frac{1}{N^3} \frac{d^2 N^2}{\Omega^2 (r^d-1)^2 } \approx  \frac{d^2}{2N \Omega^2 r^{2d} }
\ff
The result is
\f
{\cal V} =   \int d^d z \rho \left (  \frac{1}{R^2 } - 
 ( \frac{1}{\rho } \partial \rho )^2 +\frac{g'}{N^{\frac{2}{d}}} \frac{d}{d+2} r^{\prime 2} \frac{(\nabla^2 \rho )^2}{\rho^2}
+\ldots
\right )
\ff
Here we ignore total derivatives, which don't contribute to the potential energy.  
The first term is an ignorable constant.  The second term is what we want;  its variation gives the Bohmian quantum potential.  

We must  recall that in the action the ${\cal U}$ is multiplied by
$\frac{g^\prime}{g^2}$.  We choose this to be equal to
\f
\frac{g^\prime}{g^2} = \frac{\hbar^2}{8m} Z_V
\ff


The higher order terms are suppressed by powers of 
$\frac{1}{N^{\frac{2}{d}}}$.  
The result is 
\f
\boxed{
{\cal U} = -\frac{ \hbar^2}{8m}{\cal V} =
\frac{ \hbar^2}{8m} \int d^d z \rho (\frac{1}{\rho} \partial_a \rho )^2 +    O(\frac{1}{N^{\frac{2}{d}}} )}
\ff 
which we recognize as the term whose variation gives the Bohmian quantum potential. 

The leading correction is 
\f
{\cal U}^{\Delta {\cal V}} = -\frac{ \hbar^2}{8m}\Delta {\cal V} =
- \frac{1}{N^{\frac{2}{d}}} \frac{ \hbar^2 r^{\prime 2}}{8m} \int d^d z \rho (\frac{1}{\rho} \nabla^2  \rho )^2 
\ff
which contributes non-linear corrections to the Schroedinger equation.

\end{itemize}

\section{Why variety leads to Bohm}

There is a simple effective field theory argument that a term proportional to Bohm's potential
must appear.  We organize an expansion of the effective action for a particle moving stochastically in
space in a probabilistic theory , in powers of derivatives.

This effective action should have dimensions of energy, and be lorentzian or gallilean
invariant as well as parity even.   It can depend on a probability 
density for the position of the particle, $z^a (t)$ , which is $\rho (z, t)$ and spatial derivatives of $\rho$,  notated $\partial_a \rho$. It could also
depend on the momentum, $p_a$, which  we have seen may be re-expressed as
$\partial_a S$.    Note
that $\rho $ is the only density in the problem.  The only place that
derivatives of $\rho$ can then appear is in functions of
\f
\frac{\partial_a \rho }{\rho }
\ff

Thus the general form is
\f
< U> = \int d^d z \rho (z)   F [ S, \frac{\partial_a \rho}{\rho},
\partial_a S ,   \dot{S} ]
\ff

where $F$ is a scalar functional on the configuration space.  
Through second order in derivatives we have only,    
\f
< U> = \int d^d z \rho (z) \left (   \alpha \dot{S} + \beta \partial_a S \partial_b S  h^{ab}  
+\gamma  \frac{\partial_a \rho}{\rho}
\frac{\partial_b \rho}{\rho}   h^{ab}     + \dots
\right )
\ff

The main failure mode of this kind of consideration is that the signs may
not work out for the energy to be bounded from below; in
particular if $\beta <0 $ the equations give us a description of classical diffusion.

The opposite sign leads to Schrodinger quatum mechanics, as se
are about to see.

\subsection{Emergence of quantum mechanics}

We can now put all three terms together, 
\f
{\cal S}^{III} =\int d^dz    \rho ( z)   \left [   \dot{S} 
-g { \partial_a S} h^{ab}  { \partial_b S}    - 
\frac{ g^\prime}{g^2}  h^{ab}
\frac{\partial_a \rho }{\rho}   \frac{\partial_b \rho }{\rho}
\right ]  + O(\frac{1}{N^{\frac{2}{d}}} )
\label{SIII}
\ff

This has a very interesting structure.  

When we vary the action (\ref{SIII}) by 
the probability density, $\rho$, you get the 
real part of the Schrodinger
equation (\ref{rSch}) , 
\f
\dot{S} = \frac{1}{2m} (\partial_i S)^2 + \frac{ g^\prime}{g^2}  \frac{\nabla \sqrt{\rho} }{\sqrt{\rho}}
\label{rSch}
\ff
whereas varying the phase, $S$ verifies the law of current
conservation, (\ref{pc}).

The wave function 
\f
\psi = \sqrt{\rho} e^{\frac{\imath}{\hbar}S} 
\ff
then satisfies the Schrodinger equation,
\f
\imath \hbar \frac{d \psi}{dt}= \left [
-\frac{\hbar^2}{2m} \nabla^2 
\right ] \psi
\ff

This enforces our constants to be
\f
\frac{g'}{g^2} = \frac{\hbar^2 }{8 m}  Z_V
\ff

\section{Looking ahead}

\subsection{The interpretation}

Bryce de Witt used to like to say that we should let
the mathematical structure of a theory dictate its
interpretation.  Whatever the merits, generally, of that advise,
this would seem to be a good case for that practice.   The reason is that the quantum mechanics is revealed to be just
a first order phenomenon in an expansion of a small
parameter around the classical Hamilton Jacobi
equation.

Let us start with $\hbar = 0 $.  We have an action principle
\f
{\cal S}^{III} =\int d^dz    \rho ( z)   \left [   \dot{S} 
-\frac{1}{m} { \partial_a S} h^{ab}  { \partial_b S}    \right ] 
\label{SIV}
\ff
 Varying this by the probability density $\rho$
 
 we reach the Hamilton-Jacobi equation,
 \f
  \dot{S} 
-\frac{1}{m} { \partial_a S} g^{ab}  { \partial_b S}  
 \ff
 together with probability conservation.
\f
\dot{\rho} = - \partial_a (  \rho   g^{ab} \partial_b S )  
\ff

The presence of the conservation law doesn't give
us room for varying the interpretation at $\hbar =0$.
And neither does it when we turn on 
the potential energy  by increasing $ g^\prime \approx \hbar^2$. 

The theory at $\hbar=0$  
has a straightforward interpretation.  We have a probabilistic 
description of the motion of a single particle which follows the
gradients of the Hamilton-Jacobi equation.   $\rho (z,t )$
is an evolving functional that gives probabilities that
the particle can be detected at different points in the emergent
space coordinatized by the $z^a_I$.

The trajectories of the emergent particles are guided by a 
pilot wave, which is the Hamilton-Jacobi function.  Thus even
at $\hbar=0$ we have a wave particle duality?

Where did that come from?   Its been there all along-prior to
$\hbar$, prior to quantum theory, reflected in the 
Poisson bracket  (\ref{uncan}).

Furthermore we understand that theory as an emergent description of a fundamental theory of events and their
causal relations.

Now we turn on $\hbar$.  {\it The interpretation doesn't
change.}   It can't as it was already complete at $\hbar=0$.

 We still have a fundamental
theory of events, causal relations, energy and momentum.
We still have an emergent coarse grained  description, in 
terms of emergent spacetime on which there are excitations
which behave like the particles of quantum mechanics.
All that changes is that new, non-local forces are introduced,
given by the variety potential.

\subsection{Summary}

We have given a new formulation of the causal theory of views which improves on earlier versions in several ways.

The fundamental Hamiltonian is simpler and both the kinetic
and potential energy terms express directly the idea of
increasing variety, or diversity of causal neighbours, or views.
The potential energy measures the diversity of views of events
acausally related to each other, whereas the kinetic energy
measures the change in views between causally closely related events.   This makes the derivation of quantum mechanics much simpler, and it may also be hoped that this is a step towards a
specially covariant theory.

We also can mention that we have accomplished here something many approaches to quantum gravity aspire to: the emergence of space from a theory without space.

This program raises some very interesting issues
related to the reality of time\cite{present} and the physics of
the brain, which are addressed in some of the papers 
indicated\cite{QQ}.

 \subsection{Celestial spheres as views}
 
 In the near future we hope to be back to present a fully
 relativistic version of the present theory.  That will likely
 incorporate the insights of recent work 
 on celestial spheres\cite{celestial}
 as sites of clarity for understanding the asymptotic limits
 of Minkowski space.  Our expectation is
 that celestial spheres are simply fully relativistic views.  
In this construction, an incoming massless quanta is represented  by a point on the $S^2$ which gives the direction from which the quanta came.  The label on the point then gives the energy.
  The group of conformal transformations is $SL(2,C)$
  which gives a representation of the lorentz group.
 
In  the relativistic case we will also make  use of the notion of  best matching from the work of Julian Barbour and collaborators\cite{best}.
  
Finally, an obvious approach to try is to copy the form of the
action of interacting relativistic particles 
used in relative locality\cite{rl1,rl2}.


\section*{ACKNOWLEDGEMENTS}

I would like to thank Stephon Alexander, 
Julian Barbour,  Herbert Bernstein,  Giovanni-Amelino-Camelia
Saint Clair Cemin,  Marina Cortes, 
Laurent Freidel,  Lucien Hardy, Ding Jai,  Jaron Lanier,   
Andrew Liddle, Joao Magueijo, Robert Spekkens, 
Roberto Mangabeira Unger,
Donna Moylan, Antony Valentini and Clelia Verde for encouragement and 
support for following
my own compass.   I want to thank also my colleagues at
PI for keeping the faith.   

This research was supported in part by Perimeter Institute for Theoretical Physics. Research at Perimeter Institute is supported by the Government of Canada through Industry Canada and by the Province of Ontario through the Ministry of Research and Innovation. This research was also partly supported by grants from NSERC, FQXi and the John Templeton Foundation.

\end{document}